\documentclass{amsart}
\usepackage{amsmath,amssymb}

\newtheorem{theorem}{Theorem}[section]

\numberwithin{equation}{section}

\def\DJ{{\hbox{D\kern-.8em\raise.15ex\hbox{--}\kern.35em}}}
\def\DJo{$\;$\kern-.4em
    \hbox{D\kern-.8em\raise.15ex\hbox{--}\kern.35em okovi\'c}}

\def\al{{\alpha}}
\def\be{{\beta}}
\def\ga{{\gamma}}
\def\de{{\delta}}
\def\sig{{\sigma}}
\def\ve{{\varepsilon}}
\def\ze{{\zeta}}
\def\vf{{\varphi}}
\def\la{{\lambda}}
\def\bR{{\mbox{\bf R}}}

\def\bC{{\mbox{\bf C}}}

\def\pF{{\mathcal{F}}}
\def\pH{{\mathcal{H}}}

\def\tr{{\rm tr\;}}

\def\vek#1{|#1\rangle}
\def\kov#1{\langle#1|}

\begin{document}

\title[The checkerboard family]{The checkerboard family of  
entangled states of two qutrits}

\author {Dragomir \v{Z}. \DJo}
\address{Department of Pure Mathematics, University of Waterloo,
Waterloo, Ontario, N2L 3G1, Canada}
\email{djokovic@uwaterloo.ca}

\date{}

\begin{abstract}
By modifying the method of Bru{\ss} and Peres, we construct two new
families of entangled two qutrit states. For all density matrices 
$\rho$ in these families we have $\rho_{ij}=0$ for $i+j$ odd.
The first family depends on 27 independent real parameters and
includes both PPT and NPT states. The second family consists of PPT 
entangled states. The number of independent real parameters of this
family is $\ge11$.
\end{abstract}

\maketitle 

\section{Introduction}

Let $\pH=\pH_A\otimes\pH_B$ be the Hilbert space for the quantum system 
consisting of two parties, A and B. A {\em product state} is a tensor product $\rho_A\otimes\rho_B$ of the states $\rho_A$ and $\rho_B$ of the first and second party, respectively. A state $\rho$ is 
{\em separable} if it can be written as a convex linear combination of 
product states. We say that a state $\rho$ is {\em entangled} if 
$\rho$ is not separable. It is PPT if its partial transpose 
$\rho^\Gamma=(1\otimes T)(\rho)$ is a positive semidefinite operator.
Otherwise it is NPT.

The Peres-Horodecki criterion \cite{AP,MH2} settles the separability 
problem for $2\otimes2$ and $2\otimes3$ bipartite quantum systems,
but in general it provides only a necessary condition for separability.
Among bipartite systems, the $3\otimes3$ case plays a special role 
since it is the smallest case for which this criterion of separability 
is not sufficient. There are many papers, such as 
\cite{BB,CB,BP,DD,KH,PH,MH4,WK}, where entangled states 
of two qutrits or families of such states with special properties 
have been constructed and explored.

In their short note \cite{BP} Bru{\ss} and Peres constructed two 
families of PPT entangled states for two qutrits. 
Their first family depends on 5 independent real parameters and 
consists of states which are fixed under partial transposition, 
with respect to the chosen orthonormal basis of $\pH_B$. 
(We do not consider at all their second family.) 
We shall extend this family to a much richer family, $\pF'$, which 
depends on at least 11 independent real parameters 
(see Theorem \ref{Teor-2} and the subsequent remark). 
The states $\rho$ in our family are also PPT entangled and fixed 
under the partial transposition.

We begin with the four nonzero complex vectors
\begin{eqnarray*}
\vek{v_1} &=& |a,0,b;0,c,0;d,0,e\rangle, \\
\vek{v_2} &=& |0,f,0;g,0,h;0,i,0\rangle, \\
\vek{v_3} &=& |j,0,k;0,l,0;m,0,n\rangle, \\
\vek{v_4} &=& |0,p,0;q,0,r;0,s,0\rangle,
\end{eqnarray*}
where the components are listed in the order $00,10,20;01,\ldots$.
Note that $\vek{v_1}$ and $\vek{v_3}$ are orthogonal to $\vek{v_2}$ and $\vek{v_4}$, but
$\vek{v_1}$ and $\vek{v_3}$, and also $\vek{v_2}$ and $\vek{v_4}$, 
in general are not orthogonal to each other. For comparison note that 
the four vectors used in \cite{BP} are pairwise orthogonal by 
construction. By using the above vectors, we construct the normalized 
state $\rho$, i.e., with $\tr(\rho)=1$,
\begin{equation} \label{BP}
\rho=\frac 1 N \sum_{j=1}^4 \vek{v_j}\kov{v_j}, \quad 
N=\sum_{j=1}^4 \langle v_j|v_j \rangle.
\end{equation}

In the next section we prove that this $\rho$ provides a family, $\pF$, 
of entangled states of two qutrits. As we show later, $\pF$
depends on 27 independent real parameters. We refer to it as 
the {\em checkerboard family} since the corresponding density matrices
$\rho$ have the ``checkerboard pattern'', i.e., $\rho_{ij}=0$ for 
$i+j$ odd.

In section \ref{force} we construct a subfamily $\pF'\subset\pF$ 
consisting of states $\rho$ which are fixed under partial transposition. Thus, we obtain a new family of PPT entangled states of 
two qutrits. We will show later that $\pF'$ depends on at least 11 independent real parameters. It contains generically the first 
Bru{\ss}-Peres family. 
In section \ref{param} we prove the two claims on the number of 
independent real parameters. In the last section we summarize and 
discuss our results.

\section{Proving that $\rho$ is entangled} \label{ent}

As in \cite{BP}, we use the range criterion of P. Horodecki \cite{PH} 
to prove that $\rho$ is entangled in the generic case. It suffices to 
show that the subspace spanned by the $\vek{v_j}$ contains no nonzero 
tensor product of two vectors.

Assume that 
$$ \vek{\al,\be,\ga} \otimes \vek{\de,\ve,\ze}=\sum A_j\vek{v_j}
\ne0. $$
This equation is equivalent to the system of two matrix equations:
\begin{eqnarray} \label{prve}
\left[ \begin{array}{cc} a&j\\b&k\\c&l\\d&m\\e&n \end{array} \right]
 \cdot \left[ \begin{array}{c} A_1\\A_3 \end{array} \right] =
\left[ \begin{array}{c} 
\al\de \\ \ga\de \\ \be\ve \\ \al\ze \\ \ga\ze \end{array} \right] , \\
\label{druge}
\left[ \begin{array}{cc} f&p\\g&q\\h&r\\i&s \end{array} \right]
 \cdot \left[ \begin{array}{c} A_2\\A_4 \end{array} \right] =
\left[ \begin{array}{c} 
\be\de \\ \al\ve \\ \ga\ve \\ \be\ze \end{array} \right].
\end{eqnarray}

Since $(\al\de)\cdot(\ga\ze)=(\ga\de)\cdot(\al\ze)$, Eq. (\ref{prve}) 
implies that
$$ (aA_1+jA_3)(eA_1+nA_3)=(bA_1+kA_3)(dA_1+mA_3), $$
i.e.,
\begin{equation} \label{jedan}
F(A_1,A_3):=(ae-bd)A_1^2+(an+ej-bm-dk)A_1A_3+(jn-km)A_3^2=0.
\end{equation} 
Similarly, from Eqs. (\ref{prve}) and (\ref{druge}) we obtain
that
\begin{eqnarray*}
(fA_2+pA_4)(gA_2+qA_4)&=&(aA_1+jA_3)(cA_1+lA_3), \\
(hA_2+rA_4)(iA_2+sA_4)&=&(eA_1+nA_3)(cA_1+lA_3), \\
(gA_2+qA_4)(iA_2+sA_4)&=&(dA_1+mA_3)(cA_1+lA_3), \\
(fA_2+pA_4)(hA_2+rA_4)&=&(bA_1+kA_3)(cA_1+lA_3).
\end{eqnarray*}
These four equations can be written in matrix form
$$ \left[ \begin{array}{cccc} 
fg & fq+gp & pq & (aA_1+jA_3)(cA_1+lA_3) \\
hi & hs+ir & rs & (eA_1+nA_3)(cA_1+lA_3) \\
fh & fr+hp & pr & (dA_1+mA_3)(cA_1+lA_3) \\
gi & gs+iq & qs & (bA_1+kA_3)(cA_1+lA_3)
\end{array} \right] \cdot \left[ 
\begin{array}{c} A_2^2 \\ A_2A_4 \\ A_4^2 \\ -1 \end{array} \right] 
= 0. $$
Hence this $4\times4$ matrix must be singular, which gives
$$ (fs-ip)(gr-hq)(cA_1+lA_3)(\la A_1+\mu A_3)=0, $$
where
\begin{eqnarray} \label{dva}
\la &=& a(hs-ir)+b(iq-gs)+d(fr-hp)+e(gp-fq), \\ \label{tri}
\mu &=& f(mr-nq)+g(np-ks)+h(js-mp)+i(kq-jr).
\end{eqnarray}
We now assume that
\begin{equation} \label{Uslov-1}
(fs-ip)(gr-hq)F(l,-c)F(\mu,-\la)\ne0.
\end{equation}
Then $cA_1+lA_3=0$ or $\la A_1+\mu A_3=0$. Because
$F(l,-c)\cdot F(\mu,-\la)\ne0,$ neither of these two linear equations has a common nontrivial solution with Eq. (\ref{jedan}).
Consequently, we must have $A_1=A_3=0$. 
The four equations below Eq. (\ref {jedan}) now imply that 
$A_2=A_4=0$, which is a contradiction.

Thus the following theorem holds.
\begin{theorem} \label{Teor-1}
Let $a,b,\ldots,n,p,q,r,s$ be $18$ complex numbers subject to the condition (\ref{Uslov-1}), where $\la$ and $\mu$ are defined by the Eqs. (\ref{dva}) and (\ref{tri}), and $F$ by Eq. (\ref{jedan}). Then the state $\rho$ given by (\ref{BP}) is entangled. 
\end{theorem}

We denote by $\pF$ the family of entangled states $\rho$ provided by this theorem. We shall see later that $\pF$ depends only on 27 independent real parameters. 

It is easy to test whether a given state $\rho\in\pF$ is PPT. One just 
has to compute $\rho^\Gamma$ and apply to it the well known criterion 
for a hermitian matrix to be positive semidefinite. An interesting open 
question is to decide whether there is a state $\rho\in\pF$ such that 
$\rho^\Gamma$ is positive definite.

\section{Two examples}
\label{primeri}

Note that while each $\rho\in\pF$ has rank at most 4, its partial 
transpose $\rho^\Gamma$ may be nonsingular. We give here two examples 
of distillable states $\rho\in\pF$ which are of different types. In both examples $\rho^\Gamma$ is nonsingular. 

The reduction criterion provides also a necessary condition for 
separability of bipartite states $\rho$. It requires that the
matrices $\rho_A\otimes1-\rho$ and $1\otimes\rho_B-\rho$ be
positive semidefinite, where $\rho_A$ and $\rho_B$ are the two 
reduced density matrices of $\rho$. If $\rho$ violates the
reduction criterion then it is entangled. Furthermore, in that case
$\rho$ is also distillable \cite{MH1}. 

Our first example $\rho_1$ violates the reduction criterion and so 
it is distillable. The components $a,b,\ldots,s$ are chosen as follows:
$a=f=k=n=q=s=0$, $g=p=m=1$, $j=l=-1$, $h=i={\bf i}$, $e=1-{\bf i}$,  
$b=d=-1+{\bf i}$, $c=r=-1-{\bf i}$, where ${\bf i}$ is the imaginary unit. A computation shows that $\det(\rho_1^\Gamma)=2^6\cdot7/17^9$ and that $\rho_1^\Gamma$ has exactly two negative eigenvalues. We recall from \cite{MH3} that all distillable states are NPT. 
The density matrix is
\begin{equation*}
\rho_1=\frac{1}{17} \left[ \begin{array}{ccccccccc}
1 & 0 & -1 & 0 & 1 & 0 & 0 & 0 & 0 \\
0 & 1 & 0 & 0 & 0 & -{\bf i} & 0 & -{\bf i} & 0 \\
-1 & 0 & 3 & 0 & -1-2{\bf i} & 0 & 2 & 0 & -2 \\
0 & 0 & 0 & 1 & 0 & 0 & 0 & -1+{\bf i} & 0 \\
1 & 0 & -1+2{\bf i} & 0 & 3 & 0 & 2{\bf i} & 0 & -2{\bf i} \\
0 & {\bf i} & 0 & 0 & 0 & 1 & 0 & 1 & 0 \\
0 & 0 & 2 & 0 & -2{\bf i} & 0 & 2 & 0 & -2 \\
0 & {\bf i} & 0 & -1-{\bf i} & 0 & 1 & 0 & 3 & 0 \\
0 & 0 & -2 & 0 & 2{\bf i} & 0 & -2 & 0 & 2
\end{array} \right].
\end{equation*}

Our second example is a state $\rho_2\in\pF$ which is distillable but satisfies the reduction criterion:
$a=b=c=f=j=m=p=r=1$, $n=0$, $e=-1$, $q=s={\bf i}$, 
$g=h=-{\bf i}$, $d=1+{\bf i}$, $i=-1-{\bf i}$, $k=l=-1+{\bf i}$. 
The density matrix is
\begin{equation*}
\rho_2=\frac{1}{21} \left[ \begin{array}{ccccccccc}
2 & 0 & 2-{\bf i} & 0 & -{\bf i} & 0 & -{\bf i} & 0 & -1 \\
0 & 2 & 0 & 0 & 0 & 2+{\bf i} & 0 & 1+{\bf i} & 0 \\
2+{\bf i} & 0 & 3 & 0 & 0 & 0 & 0 & 0 & -1-{\bf i} \\
0 & 0 & 0 & 2 & 0 & -1 & 0 & 1+{\bf i} & 0 \\
{\bf i} & 0 & 0 & 0 & 3 & 0 & 3 & 0 & -1 \\
0 & 2-{\bf i} & 0 & -1 & 0 & 3 & 0 & 1 & 0 \\
{\bf i} & 0 & 0 & 0 & 3 & 0 & 3 & 0 & -1 \\
0 & 1-{\bf i} & 0 & 1-{\bf i} & 0 & 1 & 0 & 2 & 0 \\
-1 & 0 & -1+{\bf i} & 0 & -1 & 0 & -1 & 0 & 1
\end{array} \right].
\end{equation*}
The vector $\vek{\psi}=\vek{\vf_1}\otimes\vek{\vf_2}+
\vek{\vf_3}\otimes\vek{\vf_4}$, where 
$$ \vek{\vf_1}=\left[ \begin{array}{c} -{\bf i}\\ 1-{\bf i}\\ 1-{\bf i} \end{array} \right],\quad 
\vek{\vf_2}=\left[ \begin{array}{c} 1-{\bf i}\\ -1+{\bf i}\\ 1 \end{array} \right],\quad 
\vek{\vf_3}=\left[ \begin{array}{c} 1-{\bf i}\\ {\bf i}\\ 0 \end{array} \right],\quad 
\vek{\vf_4}=\left[ \begin{array}{c} -1\\ -{\bf i}\\ 1 \end{array} \right], $$
has Schmidt rank 2, and a computation shows that 
$\kov{\psi}\rho_2^\Gamma\vek{\psi}=-5/21$ is negative. Therefore 
$\rho_2$ is 1-distillable. We have 
$\det(\rho_2^\Gamma)=2\cdot11\cdot19/(3^7 7^9)$ and exactly two eigenvalues of $\rho_2^\Gamma$ are negative.

\section{Forcing $\rho$ to be fixed by the partial transposition}
\label{force}

Recall that all states $\rho$ of the first Bru{\ss}-Peres 
family are not only PPT but also satisfy $\rho^\Gamma=\rho$. 
We shall also impose this stronger condition.

In this section we assume that $\rho\in\pF$ and that the conditions of Theorem \ref{Teor-1} hold. The equality $\rho^\Gamma=\rho$ holds
if and only if the following eight conditions are satisfied. 
The first five conditions are 
\begin{eqnarray}
\label{j1} fg^*+pq^* &=& ca^*+lj^*, \\
\label{j2} ig^*+sq^* &=& cd^*+lm^*, \\
\label{j3} fh^*+pr^* &=& cb^*+lk^*, \\
\label{j4} ih^*+sr^* &=& ce^*+ln^*, \\
\label{j5} ae^*+jn^* &=& db^*+mk^*,
\end{eqnarray}
and the remaining three are
\begin{equation} \label{uslovi}
\Im(ad^*+jm^*)=\Im(be^*+kn^*)=\Im(fi^*+ps^*)=0, 
\end{equation}
where $\Im(z)$ denotes the imaginary part of the complex number $z$.
In the following analysis we shall assume that all denominators in our 
formulae are nonzero. 

Since $fs-ip\ne0$, the equations (\ref{j1}) and (\ref{j2}) can be 
solved for $g$ and $q$, and the equations (\ref{j3}) and (\ref{j4}) for 
$h$ and $r$:
\begin{eqnarray} \label{jg}
g &=& \frac {s^*(ac^*+jl^*)-p^*(dc^*+ml^*)} {(fs-ip)^*}, \\ \label{jq}
q &=& \frac {f^*(dc^*+ml^*)-i^*(ac^*+jl^*)} {(fs-ip)^*}, \\ \label{jh}
h &=& \frac {s^*(bc^*+kl^*)-p^*(ec^*+nl^*)} {(fs-ip)^*}, \\ \label{jr}
r &=& \frac {f^*(ec^*+nl^*)-i^*(bc^*+kl^*)} {(fs-ip)^*}. 
\end{eqnarray}

{}From Eqs. (\ref{uslovi}) we obtain that 
\begin{equation} \label{jdei}
d=(x-mj^*)/a^*, \quad e=(y-nk^*)/b^*, \quad i=(t-sp^*)/f^*,
\end{equation}
where $x,y,$ and $t$ are real numbers.
By plugging in the above expressions for $d$ and $e$ into Eq. 
(\ref{j5}) and solving it for $n$, we obtain that
\begin{equation} \label{jn}
n=\frac {|a|^2y-|b|^2x-m^*b^*(ak-bj)} {a(ak-bj)^*}.
\end{equation}

To summarize, we have the following result.

\begin{theorem} \label{Teor-2}
Let $t,x,y$ be real and $a,b,c,f,j,k,l,m,p,s$ complex parameters and 
define $g,q,h,r,d,e,i,n$ by the formulae (4.7-12). Assume that 
$$ abf(fs-ip)(gr-hq)(ak-bj)F(l,-c)F(\mu,-\la)\ne0, $$
where $\la$ and $\mu$ are defined by Eqs. (\ref{dva}) and (\ref{tri}) 
and $F$  by Eq. (\ref{jedan}).
Then the state $\rho$ given by (\ref{BP}) is entangled and satisfies
$\rho^\Gamma=\rho$. In particular, $\rho$ is a PPT state.
\end{theorem}

We denote by $\pF'$ the subfamily of $\pF$ singled out by this theorem. 
Since all states $\rho\in\pF'$ satisfy $\rho^\Gamma=\rho$, it is
clear that $\pF'$ does not contain the second Bru{\ss}-Peres family.

On the other hand, we remark that $\pF'$ contains generically their first family. To prove this we set
$$ k=y=0,\, j=c^*,\, l=-a^*,\, m=x/c,\, p=tcf^*/xa,\, s=xa^*/fc^*. $$
Then the above formulae for $g,q,h,r,d,e,i,n$ give
$$ d=e=i=n=r=0,\, g=tfc^*,\, q=-f^*,\, h=bc^*/f^*, $$
and from the Eq. (\ref{jedan}) we obtain that
$$ F(l,-c)=-xba^*, \quad F(\mu,-\la)=\frac {x(bc^*)^3} {|acf^2|^2}
\cdot (x|a|^2-t|f|^2)^2. $$ 
By assuming that also $a,b,c,f$ are real and that 
$abcfx(x|a|^2-t|f|^2)\ne0$, we obtain generically the first 
Bru{\ss}-Peres family. Explicitly, our parameters $a,b,c,f,t,x$
correspond to $m,ac/n,n,a,b/an,adn/m$, respectively, in the
Bru{\ss}-Peres family. If we allow $a,b,c,f$ to remain complex, 
the number of independent real parameters increases from 5 to 7.

\section{Counting independent real parameters} \label{param}

We consider first the family $\pF$ provided by 
Theorem \ref{Teor-1}. The state $\rho$ depends on 36 real 
parameters (the real and imaginary parts of the 18 complex 
parameters). However, we claim that there are only 27 independent 
such parameters. In order to prove this claim, it is convenient 
to drop the normalization factor $1/N$ and show that the modified 
family $\Phi:(a,b,\ldots,s)\mapsto N\rho$ depends on  
28 independent parameters. 

Since the rank of $d\Phi$ can take only finitely many values, 
there exists a point $\omega_0=(a_0,b_0,\ldots,s_0)$ in the domain 
of $\Phi$ at which the rank of $d\Phi$ takes the maximal value, 
say $\delta$. By continuity of $d\Phi$, its rank has to be 
$\delta$ in some small neighborhood $U$ of the point $\omega_0$. 
Now the Rank Theorem \cite[Theorem 7.1]{WB} implies that the image $\Phi(U)\subseteq\pF$ is a manifold of dimension $\delta$. 
Therefore the number of independent real parameters of $\pF$ 
is at least $\delta$.

We shall first find an upper bound for $\delta$.
Due to the checkerboard pattern of $\rho$, by simultaneous permutation of rows and columns of the matrix $N\rho$ we obtain the direct sum $\rho' \oplus \rho''$ where
$$
\rho'=\left[ \begin{array}{cccc}
gg^*+qq^* & gf^*+qp^* & gi^*+qs^* & gh^*+qr^* \\
fg^*+pq^* & ff^*+pp^* & fi^*+ps^* & fh^*+pr^* \\
ig^*+sq^* & if^*+sp^* & ii^*+ss^* & ih^*+sr^* \\
hg^*+rq^* & hf^*+rp^* & hi^*+rs^* & hh^*+rr^* 
\end{array} \right]
$$
and
$$
\rho''=\left[ \begin{array}{ccccc}
aa^*+jj^* & ad^*+jm^* & ac^*+jl^* & ab^*+jk^* & ae^*+jn^* \\
da^*+mj^* & dd^*+mm^* & dc^*+ml^* & db^*+mk^* & de^*+mn^* \\
ca^*+lj^* & cd^*+lm^* & cc^*+ll^* & cb^*+lk^* & ce^*+ln^* \\
ba^*+kj^* & bd^*+km^* & bc^*+kl^* & bb^*+kk^* & be^*+kn^* \\
ea^*+nj^* & ed^*+nm^* & ec^*+nl^* & eb^*+nk^* & ee^*+nn^*
\end{array} \right].
$$
Both $\rho'$ and $\rho''$ have rank 2. For instance, the nullspace of 
$\rho'$ is spanned by the columns of the matrix
$$
\left[ \begin{array}{cccc}
(fs-ip)^* & (fr-hp)^* \\
(iq-sg)^* & (hq-gr)^* \\
(gp-fq)^* & 0 \\
0 & (gp-fq)^* 
\end{array} \right].
$$
It follows that the rank of $d\Phi$ is the same as that of the map
$$ \Psi: (a,b,\ldots,s) \mapsto \sig' \oplus \sig'', $$
where $\sig'$ resp. $\sig''$ is the submatrix of $\rho'$ resp. 
$\rho''$ consisting of the first two columns. Since $\rho'$ and 
$\rho''$ are hermitian, the image of $\Psi$ is contained in a 
28-dimensional real vector space. Consequently, $\Psi$ depends on at
most 28 independent real parameters and $\delta\le28$.

On the other hand, a computation shows that the rank of $d\Psi$ at 
the point used to define the state $\rho_2$ (see Sect. \ref{primeri}) 
is indeed 28. We conclude that $\delta=28$ and our claim follows.

Next we consider the family $\pF'$ constructed in Theorem \ref{Teor-2}. 
Our second claim is that $\pF'$ depends on at least 11 independent 
real parameters. In order to prove this claim, we again drop the 
normalization factor $1/N$ and it suffices to show that the 
modified family 
$$ \Lambda:(t,x,y,a,b,c,f,j,k,l,m,p,s)\to N\rho $$ 
depends on at least 12 independent parameters. 

In total we have 23 real parameters: $t,x,y$ and the real and 
imaginary parts of $a,b,c,f,j,k,l,m,p,s$. The domain of $\Lambda$ 
is an open subset of $\bR^3 \times \bC^{10}$, i.e., $\bR^{23}$. 
The matrix entries $\rho_{ij}$ are complex-valued rational functions 
of these parameters. Since the rank of the Jacobian matrix of 
$\Lambda$ at the point: $t=1$, $x=l=0$, $y=-1$,
$c=s={\bf i}$, $a=j=-{\bf i}$, $b=f=1-{\bf i}$, $p=1+{\bf i}$,
$k=m=-1+{\bf i}$ is 12, $\Lambda$ indeed depends on at least 12
independent real parameters. Hence our second claim follows.

\section{Discussion} \label{zakljucak}

We have shown that the states $\rho$ of two qutrits given by Eq. 
(\ref{BP}) are generically entangled. The precise conditions are 
spelled out in Theorem \ref{Teor-1}. We denote by $\pF$ the family 
of states $\rho$ as in Eq. (\ref{BP}) satisfying the conditions of 
this theorem. Although it is not obvious, one can verify that 
each $\rho\in\pF$ has rank 4, and so its eigenvalue 0 has 
multiplicity 5. Just as in \cite{BP}, for $\rho\in\pF$ we have 
$\rho_{ij}=0$ whenever $i+j$ is odd. Formally, $\pF$ depends on 
36 real parameters but only 27 of them are independent. 
The proof of this fact is given in section \ref{param}. 

Let us say here that a state $\rho\in\pF$ is {\em generic} if 
$\rho^\Gamma$ is nonsingular. The family $\pF$ contains a variety of
distillable states. In Section \ref{primeri} we give two such 
examples, $\rho_1$ and $\rho_2$, which are also generic. 
The former violates the reduction criterion of separability while 
the latter does not. We show explicitly that $\rho_2$ is 1-distillable.
We point out that the first nontrivial example of a distillable 
bipartite state which satisfies the reduction criterion was given 
in \cite{PS}.

Next we have singled out an explicit subfamily $\pF'\subset\pF$ 
consisting of PPT-states. In fact the states $\rho\in\pF'$ are fixed
under partial transposition, i.e., we have $\rho^\Gamma=\rho$ for
all $\rho\in\pF'$. This subfamily was selected so to contain almost 
all members of the first Bru{\ss}-Peres family. Our family $\pF'$ 
is much richer than the latter as it depends on at least 11 
independent real parameters.  

Let us mention that Bru{\ss}-Peres family was used in \cite{JV} 
to test the effectiveness of a new separability condition.
We hope that our families $\pF$ and $\pF'$ will also be useful for 
further study of entanglement and distillability problems. 
There is no doubt that our first theorem admits a generalization to
bipartite systems of qudits. It would be worthwhile to work out the 
details at least in the $4\otimes4$ case.


\begin{thebibliography}{99}

\bibitem{BB}
Baumgartner, B., Hiesmayr, B.C., and Narnhofer, H.,
Phys. Rev. A {\bf 74}, 032327 (2006). 
Phys. Lett. A 372 (2008) 2190-2195;

\bibitem{CB}
Bennett, C.H., DiVincenzo, D.P., Mor, T., Shor, P.W., Smolin, J.A.
and Terhal, B.M.,
Phys. Rev. Lett. {\bf 82}, 5385-5388 (1999). 

\bibitem{WB}
Boothby, W.M.,
An Introduction to Differentiable Manifolds and Riemannian Geometry, Academic Press, New York, 1975.

\bibitem{BP}
Bru{\ss}, D., Peres, A., Phys. Rev. A {\bf 61}, 030301(R) (2000).

\bibitem{JV}
de Vincente, J.I.,
Quantum Inf. Comput. {\bf 7}, 624 (2007).

\bibitem{DD}
DiVincenzo, D.P., Mor, T., Shor, P.W., Smolin, J.A. and Terhal, B.M.,
Comm. Math. Phys. {\bf 238}, 379-410 (2003). 

\bibitem{KH}
Ha, K.-C. and Kye, S.-H.,
J. Phys. A: Math. Gen. {\bf 38} (2005), 9039-9050.

\bibitem{MH1}
Horodecki, M. and Horodecki, P., 
Phys. Rev. A {\bf 59} 4206 (1999). 

\bibitem{MH2}
Horodecki, M., Horodecki, P. and Horodecki, R.,
Phys. Lett. A 223 (1996) 1-8. 

\bibitem{MH3}
Horodecki, M., Horodecki, P. and Horodecki, R.,
Phys. Rev. Lett. {\bf 80}, 5239-5242 (1998). 

\bibitem{MH4}
Horodecki, M., Horodecki, P. and Horodecki, R., in
{\em Quantum Information Theory: An Introduction to Basic
Theoretical Concepts and Experiments}, edited by G. Alber 
{\em et al.}, Springer Tracts in Modern Physics Vol. 173
(Springer Verlag, Berlin, 2001), pp. 151-195.

\bibitem{PH}
Horodecki, P., Phys. Lett. A 232 (1997) 333-339. 

\bibitem{WK}
Kim, W.C. and Kye, S.-H.,
Phys. Lett. A 369 (2007) 16-22.

\bibitem{AP}
Peres, A.,
Phys. Rev. Lett. {\bf 77}, 1413-1415 (1996).

\bibitem{PS}
Shor, P.W., Smolin, J.A. and Terhal, B.M.,
Phys. Rev. Lett. {\bf 86}, 2681-2684 (2001).


\end{thebibliography}
\end{document}